\documentclass{article}
\usepackage{psfig}
\newcommand{\be}{\begin{equation}}
\newcommand{\ee}{\end{equation}}
\newcommand{\ba}{\begin{array}{c}}
\newcommand{\ea}{\end{array}}
\newcommand{\bea}{\begin{eqnarray}}
\newcommand{\eea}{\end{eqnarray}}
\begin{document}
\date{\today} 
\title{Effect of the unpolarized spin state in spin-correlation measurement of two protons produced in the ${\rm{^{12}C}}(d$,$\rm{^{2}He)}$reaction}\maketitle 
\begin{center}
S. Hamieh\\[0.3cm] {\it Kernfysisch Versneller Instituut,\\ Zernikelaan 25, 9747 AA Groningen, The Netherlands.}
\end{center}
\begin{abstract}In this note we discuss the effect of the unpolarized state in the spin-correlation measurement of the $^1S_0$ two-proton state produced in ${\rm{^{12}C}}(d$,$\rm{^{2}He)}$ reaction at the KVI, Groningen.
 We show that in the presence of the unpolarized state the {\it maximal} violation of  the CHSH-Bell inequality is lower than the classical limit if the purity of the state  is less than $ \sim \verb+70%+$. In particular, for  the KVI experiment the violation of the CHSH-Bell inequality should be corrected by a factor  $\sim\verb+10%+$ from the pure $^1S_0$ state.
\end{abstract}
\newpage
\section{Introduction}
In an experiment performed at the   Kernfysisch Versneller Instituut (KVI), Groningen \cite{kvi} with the goal to test Bell inequality violation in Nuclear Physics (perhaps to be applied in quantum information physics), the experimental group, by bombarding a $^{12}\rm C$ target with 170 MeV $d$, was able to generate a singlet-spin, two-proton state coupled to unpolarized  state with $\sim 10\verb+%+$ contribution. In this paper we will analyze the experimental results of this experiment and we will show that the effect of the unpolarized state  is important and could not be neglected.
\section{CHSH inequalities and entanglement in a mixed\\ ensemble}
 Bell-type inequalities relating averages of four random dichotomic variables ${a}$, ${a^{\prime}}$ and ${b}$, ${b^{\prime}}$ representing measurements of spin in directions {\bf $\hat{a}$}, {\bf $\hat{a^{\prime}}$} and {\bf $\hat{b}$}, {\bf $\hat{b^{\prime}}$}. The Clauser, Horne, Shimony and Holth (CHSH) \cite{Clau69} form of Bell-type inequalities for spin $1/2$ case could be written in this form  \be  |E(\phi_1,\phi_1^{'},\phi_2,\phi_2^{'})|=|E(\phi_1,\phi_2)+E(\phi_1,\phi_2^{'})+E(\phi_1^{'},\phi_2)-E(\phi_1^{'},\phi_2^{'})|\leq 2,\ee where $\phi_i$ is the analyzer angular setting for the $i^{th}$ particles ($i=1$, $2$) and $E(\phi_i,\phi_j)$ is the correlation function defined as \be E(\phi_i,\phi_j)={N_{++}+N_{--}-N_{+-}-N_{-+} \over N_{total}}\,.\ee In quantum-theory language  the CHSH operator corresponding to the  CHSH inequality is represented by an operator \be  {\cal{B}}= \hat{a} {\bf \cdot\sigma}\otimes (\hat{b}+\hat{b^{\prime}}) {\bf\cdot\sigma}+ \hat{a^{\prime}} {\bf \cdot\sigma}\otimes (\hat{b}-\hat{b^{\prime}}) {\bf\cdot\sigma}\,,\ee acting in Hilbert space ${\cal{H_{\rm A}}\otimes\cal{H_{\rm B}}}$ in $2\otimes 2$ dimension. The correlation function is given by the  mean value of the operator {\bf $\hat{a}\sigma\otimes \hat{b}\sigma$}. For  a pure state this correlation function could be easily computed, $e.g.$ for singlet state we have
\be E(\phi_i,\phi_j)=-cos(\phi_i-\phi_j)\,.\ee For mixed state, however, the mean value should be averaged over the ensemble and therefore the CHSH inequality not a sufficient condition to test the presence of entanglement\cite{Wern89}.  Different measures of the entanglement have been proposed in the literature for mixed state\footnote{Any  measurement of the entanglement should not increase by local operation ({\it e.g.} unitary transformation) and classical communication ({\it e.g.} phone calls.), known as LOCC},  e.g. entanglement of formation, distillation, relative entropy of entanglement, negativity, etc$\dots $ Here we will use the entanglement of formation as our measure of the entanglement. 

In  a mixed ensemble any bipartite quantum state $\rho_{AB}$  can be written as: \be\label{eq3} \rho_{AB}={1\over 4}\left(I\otimes I + {\bf A\cdot\sigma}\otimes I +I \otimes {\bf P\cdot\sigma} +\sum_{i,j=1}^3D_{ij}\sigma_i\otimes\sigma_j\right )\,. \ee $\sigma_i$ are the pauli matrices, $I$ is the identity operator, ${\bf A}$ and ${\bf P}$ are vectors in ${\cal {R}}^3$. The $D_{ij}$ form a $3\times 3$ matrix called the correlation matrix $D$. In this representation of the density matrix the mean value of the CHSH-Bell operator is given by \cite{Horo95} \be \langle {\cal{B}} \rangle=\hat{a}\cdot\left[D(\hat{b}+\hat{b^{\prime}})\right]+\hat{a^{\prime}}\cdot\left[D(\hat{b}-\hat{b^{\prime}})\right]\,.\ee
Using the representation of the density matrix given in Eq. (\ref{eq3}), we characterize any bipartite quantum state $\rho_{AB}$ by
\begin{itemize}
\item The entanglement  measured by the  ``tangle'', $\tau$, of the entanglement of formation \cite{Benn96} and defined by \be\tau=\max\{\lambda_1-\lambda_2-\lambda_3-\lambda_4,0\}\,,\ee where the $\lambda$'s are the square roots of the eigenvalues, in decreasing order, of the matrix, $\rho_{AB} (\sigma_y\otimes\sigma_y\rho_{AB}^{\star}\sigma_y\otimes\sigma_y)$ and $ \rho_{AB}^{\star}$ is the complex conjugation of $\rho_{AB}$ in the computational basis $\{|++\rangle,|+-\rangle,|-+\rangle,|--\rangle\}$.
\item The maximum  amount of the CHSH-Bell violation of the state $\rho_{AB}$  \cite{Horo95}
\be \langle{\cal{B}}\rangle^{max}=2\sqrt{M(\rho_{AB})}\,.\ee $M(\rho_{AB})$ is the sum of the two larger eigenvalues of $DD^{\dag}$ \footnote{In this case the directions  $\hat{b}$ and $\hat{b^{\prime}}$ of the analyser setting are equal to $ \cos(\theta)\hat{c}_{\rm max}\pm \sin(\theta)\hat{c}_{\rm max}^{\prime}$ and the direction $\hat{a}$, $\hat{a^{\prime}}$ are equal to $D\hat{c}_{\rm max}\over ||D\hat{c}_{\rm max}||$, $D\hat{c}^{\prime}_{\rm max}\over ||D\hat{c}^{\prime}_{\rm max}||$, respectively. $\hat{c}_{\rm max}$ and $\hat{c}_{\rm max}^{\prime}$ are  two unit (not unique) and mutually orthogonal vectors in ${\cal R}^3$ that maximize the function $||D\hat{c}||^2+||D\hat{c}^{\prime}||^2$ (see Ref. \cite{Horo95} for more detail).}. \item The purity of the state that measures how far the state is from pure state \be S_L={\rm Tr}(\rho_{AB}^2)\,.\ee \end{itemize}
\section{Analysis of the experimental data of the KVI experiment} 
The spin state of the two protons produced in the ${\rm{^{12}C}}(d$,$\rm{^{2}He)}$ reaction at $E_d=170$ MeV at KVI \cite{kvi} is a singlet state mixed with the unpolarized (random contamination) state with $\gamma$ ($0\leq\gamma\leq 1$) controlling the degree of  mixing (the details of the experimental setup and analysis  method of the ($d$,$^2$He) reaction were described in detail in  Ref. \cite{kvi}). Given all that,  we can write the density matrix of such state as \be \rho_{W}=(1-\gamma){I\over 4}+\gamma|\Psi^{-}\rangle\langle\Psi^{-}|\,\ee which interpolates between the unpolarized state $I/4$ and singlet state $|\Psi^{-}\rangle=(|+-\rangle-|-+\rangle)/\sqrt(2)$. This class of states is called Werner states \cite{Wern89}. The purity of Werner states is a monotonic function of $\gamma$. Thus, in this paper we use $\gamma$ as our measure of purity. Also, for Werner state it is easy to prove using the condition noted above that \be\langle{\cal{B}}\rangle_{ Werner}^{max}=\gamma\langle{\cal{B}}\rangle_{ pure}^{max}\,.\ee \begin{figure}[tb]\centerline{ \psfig {width=10cm,figure=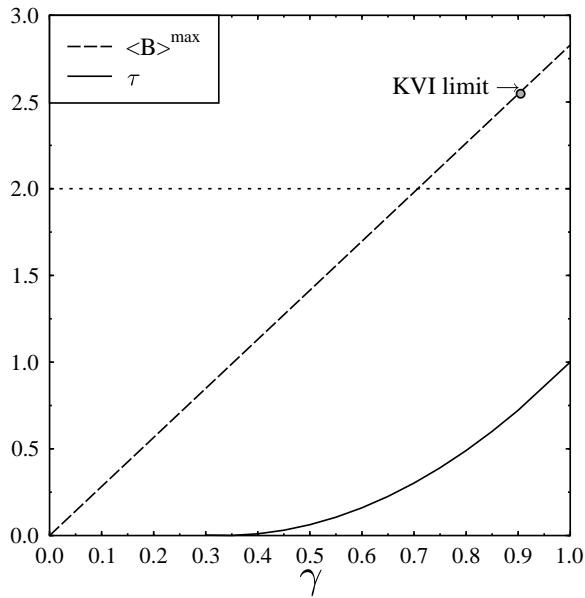}} \vspace*{-1cm}\caption{ Plot of $\langle{\cal{B}}\rangle_{Werner}^{\max}$ (dashed line) and $\tau$ (solid line) versus the purity $\gamma$. The dotted line is the Bell limit, the circle is the KVI limit for $\gamma\sim 0.9$. \protect\label{figure1}}\end{figure}

\begin{table}[t]
\caption{\label{table1}Experimental data and quantum theory predictions for a pure singlet states (case 1) and mixed Werner states (case 2) for several violating cases of the CHSH-Bell inequality according to the definition given in Eq. 1.}
\begin{center}
\begin{tabular}{l|l|l|l}\hline\hline
CHSH-Bell &   QM &QM  &Exp. Data\\
Inequality&     case 1&case 2 &   \\ \hline
$E(50^{\circ},0^{\circ},25^{\circ},75^{\circ}$)   &   2.46  &2.21& 0.67$\pm$ 2.30 \\ \hline
$E(60^{\circ},0^{\circ},30^{\circ},90^{\circ})$   &   2.60  &2.34& 1.21$\pm$ 2.42  \\ \hline
$E(70^{\circ},0^{\circ},35^{\circ},105^{\circ})$  &   2.72 &2.45 & 1.54$\pm$ 2.76   \\ \hline
$E(80^{\circ},0^{\circ},40^{\circ},120^{\circ})$  &   2.80 &2.52 & 2.11$\pm$ 2.86    \\ \hline
$E(90^{\circ},0^{\circ},45^{\circ},135^{\circ})$  &   2.83 &2.55 & 2.23$\pm$ 2.48     \\ \hline
$E(100^{\circ},0^{\circ},50^{\circ},150^{\circ})$ &   2.79&2.51  & 2.39$\pm$ 2.87      \\ \hline
$E(110^{\circ},0^{\circ}, 55^{\circ},165^{\circ})$&   2.69&2.34  & 2.58$\pm$ 2.91       \\ \hline
$E(120^{\circ},0^{\circ},60^{\circ},180^{\circ})$ &   2.50&2.25  & 2.75$\pm$ 2.95        \\ \hline
$\chi^2$ & 1.26& 0.85&\\ \hline
\end{tabular}
\end{center}
\vskip -0.8cm
\end{table}
Note that, a violation of the modified Bell-inequality does not exclude
an explanation with a hidden variable theory.
In Fig. \ref{figure1} we plot $\langle{\cal{B}}\rangle_{Werner}^{max}$ and the tangle $\tau$ versus the purity $\gamma$. As we can see in this figure the Werner state does not violate the Bell inequality if its purity $\gamma$ is less than $1/\sqrt 2 \sim 70\verb+%+$. However, the entanglement is still non-zero  in the Werner state until  $\gamma>1/3 \sim 33\verb+%+$. Therefore, some quantum correlation cannot be seen only by measuring the violation of the Bell-type inequality because some of them (Werner states) are entangled but still do not violate Bell inequality.  Note that there is a possible experimental measurement of the entanglement based on the entanglement witness \cite{Guhn02} that we think to implement  in the future experiment.

In Tab. \ref{table1} we compare the quantum theory predictions assuming a pure singlet state (case 1) and mixed Werner states (case 2) for the spin of the two detected protons for several violating cases of the CHSH-Bell inequality. The value of $\chi^2=\sum_i({R^i_{th}-R^i_{exp}\over \Delta R^i_{exp}})^2$ is given in the bottom of the table for both cases.  We have found that $\chi^2_{Werner}<\chi^2_{Singlet}$ as expected. However, we cannot judge this result as  evidence of the mixing of the singlet with the unpolarized state because the experimental data suffer from large errors.
\section{Conclusion}
In this paper we have discussed the effect of the unpolarized state in the spin correlations measurement of the $^1S_0$ two proton state produced in ${\rm{^{12}C}}(d$,$\rm{^{2}He)}$ reaction at KVI. We have shown that even a small coupling (less than $10\verb+%+$) of the pure singlet state with the unpolarized state changes dramatically the Bell-violation value. After introducing the contribution of the    unpolarized state we have found a better $\chi^2$. The experimental results are suffering from a large statistical error and therefore not conclusive
for testing Bell's inequality, but with a modified experimental setup, measurements with significantly
improved precision will become feasible.
\section*{Acknowledgments}
This work was performed as part of the research program of the {\sl Stichting voor Fundamenteel Onderzoek der Materie (FOM)} with financial support from the {\sl Nederlandse Organisatie voor Wetenschappelijk Onderzoek }.

\end{document}